\documentclass[aps,prb,groupedaddress,reprint,showpacs]{revtex4-1}
\usepackage{graphicx}
\usepackage{bm}
\usepackage{color}
\usepackage{tabularx}
\usepackage{threeparttable}
\usepackage{tablefootnote}
\usepackage{here}

\begin{document}

\title{
  Infrared-Shielding of Plasmonic Random Metasurface Constructed by Cesium-Doped Tungsten Bronze
}
\author{Tomohiro Yoshida$^{1}$}
\author{Takashi Takeuchi$^{2}$}
\author{Kazuhiro Yabana$^{2}$}

\affiliation{$^{1}$
  Department of Computer-Aided Engineering and Development,
  Sumitomo Metal Mining Co., Ltd., 3-5, Sobiraki-cho, Niihama, Ehime 792-0001, Japan
}

\affiliation{$^{2}$
  Center for Computational Sciences, University of Tsukuba, Tsukuba 305-8577, Japan
}

%%%%%%%%%%%%%%%%%%

\begin{abstract}
  The heat-shielding properties of random metasurface,
  composed of spherical or spheroidal nanoparticles with random displacements and/or random deformation,
  were theoretically investigated using
  the finite difference time domain method.
  The effective coverage was defined using the total area of nanoparticles in the metasurface,
  and the robustness of the near-infrared light reflection against randomness was investigated.
  When the effective coverage was high, the near-infrared light reflection was reduced by at least 20\% in 
  both nanoparticle arrangement and shape randomness compared to the hexagonal close-packed perfect metasurface.
  In contrast, when effective coverage was low, 
  the randomness of the nanoparticle arrangement had almost no effect on 
  the near-infrared light reflection.
  Furthermore, the near-infrared light reflection performance was improved by 
  the randomness of the nanoparticle shape.
\end{abstract}
\maketitle
%%%%%%%%%%%%%%%%%%%%%%%%%%%%%%%%%%%%
\section{INTRODUCTION}
The heat shielding glass that blocks near-infrared (NIR) light in sunlight, used in window glass, car windshields, among others, has attracted much attention in terms of energy saving~\cite{Lee,Wang,Gu,Chang}.
The heat shielding glass absorbs NIR light using the localized surface plasmon resonance (LSPR)
owing to the metal nanoparticles dispersed in the glass~\cite{Bohren}.
Several materials that induce LSPR are known; examples include Ag, Au, and Cu and their alloys~\cite{West}, 
antimony-doped tin oxide (ATO)~\cite{Nutz, Li}, 
tin-doped indium oxide (ITO)~\cite{Ederth, Kanehara, Garcia, LiShi},
Al-doped zinc oxide~\cite{Thu}, 
rare-earth hexaborides (XB $ _6 $: X indicates rare earth elements)~\cite{Schelm, Xiao, Machida, Adachi}, 
and alkali-doped tungsten bronzes (M$_x$WO$_3$: M = Na, K, Rb, Cs, and Tl)~\cite{Mamak, Mattox, Adachi2012, machida2016}.
However, while heat-shielding glass exhibits high NIR light absorption,
a part of the absorbed light is re-emitted into the room or car as heat.
Therefore, to improve the heat-shielding property, 
transparent materials that transmit visible light and reflect NIR light are required.

Plasmonic metasurfaces with uniformly shaped metal nanoparticles arranged in two dimensions are good candidates for transparent NIR reflection.
Examples include metasurfaces constructed by spherical ITO nanoparticles.~\cite{Matsui}.
Tani {\it et al.,} realized NIR light reflection by controlling the aspect ratio ($AR$) of hexagonal Ag nanoparticles~\cite{Tani:14}.
While metasurfaces are promising transparent NIR reflective material candidates, industrial applications are challenging.
Existing metasurfaces have uniform nanoparticle shapes 
and are constructed by regularly arranged nanoparticles.
The production of uniform particles can be realized using 
the hydrothermal method~\cite{C5TC01411E,Mattox}.
However, poor production efficiency and cleaning required for removing impurities after synthesis increase production costs.
For the perfect arrangement of nanoparticles,
the organic molecule coating, requiring strict control for uniform coating, is used~\cite{Doyle}.

A low-cost method for producing nanoparticles is the breakdown method~\cite{Adachi2013,Sato2012}, where a material synthesized in the solid-state reaction is milled.
Compared to the hydrothermal method, the breakdown method has industrial merits such as high yield, 
cheaper precursors, and no additional cleaning.
However, the failure to produce uniformly shaped nanoparticles makes it difficult to coat them with organic molecules uniformly.
Heat-shielding properties of resulting metasurfaces, where the particle shape and arrangement are random, have not been investigated in detail.
Although Tani {\it et al.,} investigated the NIR reflection in the random dispersed Ag nanoparticles~\cite{Tani:14},
they did not discuss comparison with the perfect and random arrangement of nanoparticles, and randomness of nanoparticle shape.

This study is intended to investigate the NIR light reflection of a random metasurface. 
As the material, we consider a metasurface constructed using Cs$_{0.33}$WO$_3$ (CsWO) that shows LSPR~\cite{Adachi2012, machida2016}. 
We evaluate the NIR light reflection when the particle shape and/or arrangement are randomized 
using the finite different time domain (FDTD) method. 
The robustness of NIR light reflection against randomness will be discussed for metasurfaces 
with different coverages.

\section{METHOD}
In this study, the light transmittance, reflectance, and 
absorptance of the random metasurfaces made of CsWO were calculated by the FDTD method.
The relative permittivity of CsWO, $\varepsilon(\omega)$, was given by the following Lorentz-Drude model,
\begin{eqnarray}
  \varepsilon(\omega)=1-\sum_{n=1}^{3}f_n\frac{\omega_p^2}{(\omega^2-l_n^2)+i\gamma_n\omega},
  \label{eq1}
\end{eqnarray}
where $\omega_p=3.907$ eV, $f_1=1.2$, $f_2=1.6$, $f_3=2.7$, $l_1=0.0$ eV, $l_2=4.91$ eV, 
$l_3=6.0$ eV, $\gamma_1=0.2$ eV, $\gamma_2=1.0$ eV, and $\gamma_3=1.5$ eV.
We determined these parameters
to reproduce the results of the first-principles calculation~\cite{Yoshio_2018}.

\begin{table*}[htb]
  \begin{tabular}{cccc}
    \hline
     & 
    \begin{tabular}{c}
      $xy$ plane \\ random \\ metasurface
    \end{tabular}
    &
    \begin{tabular}{c}
      $xyz$ \\ random \\ metasurface
    \end{tabular}
    &  
    \begin{tabular}{c}
      $xy$ plane and 
      \\ nanoparticle shape \\ random metasurface 
    \end{tabular}
    \\
    \hline
    \begin{tabular}{c}
      arrangement in \\ $xy$ plane 
    \end{tabular}
    & random & random & random \\
    \begin{tabular}{c}
      arrangement along \\ $z$ direction 
    \end{tabular}
    & uniform & random & uniform \\
    axis ratio of spheroid & uniform & uniform & random \\
    \hline
  \end{tabular}
  \caption{
    Types of randomness.
  }  
  \label{table1}
\end{table*}

First, we describe the structure of the perfect metasurface, having a uniform nanoparticle shape 
and regularly arranged nanoparticles.
CsWO nanoparticles produced by the breakdown method was reported to be nearly spherical with 
7 nm radius~\cite{machida2016}.
Therefore, we assumed that the perfect metasurface was constructed by spherical CsWO with radius of $r=7$ nm.
The CsWO nanoparticles are hexagonally close-packed on the $xy$ plane, with $z=0$.

We introduced three types of randomness into the perfect metasurface:
randomness in nanoparticle arrangement in (i) the $xy$ plane; 
(ii) the $xyz$ directions; 
(iii) $xy$ plane with particle shape randomness.
We summarized the random metasurface in Table~\ref{table1}.
Also see Fig.~S1 displayed in the Supporting Information.
To take into account randomness in $xy$ plane, it is important to employ a unit cell in the 
$xy$ plane containing a large number of nanoparticles. 
In our calculation, we use a unit cell containing 120 nanoparticles. 
We will later show that this choice is sufficient by comparing results of different random configurations. 
The unit cell has a size of $x_l \times y_l \times 100$ nm with $y_l = 3\sqrt{3}x_l/5$. 
A periodic boundary condition is imposed in the $x$ and $y$ directions, 
and a PML absorbing boundary condition in the $z$ direction.

We introduce the effective coverage ($EC$) as follows:
\begin{eqnarray}
  EC=\frac{120 \times\pi r^2}{x_l y_l}.
\end{eqnarray}
Higher $EC$ corresponds to a hexagonal close-packed structure 
with a small inter-nanoparticle gap in a perfect metasurface.
Although metasurfaces with different inter-nanoparticle gaps have been experimentally realized, 
the higher $EC$ shows higher NIR light reflection~\cite{Matsui,Tani:14,Doyle}.
To investigate how the effect of randomness differs
between high and low $EC$, we investigated $EC$ at 0.5 and 0.8.
The nearest-neighbor nanoparticle gap in the perfect metasurface is 4.8 nm for $EC=0.5$ and 
0.8 nm for $EC=0.8$.
We show the structure of the perfect metasurface in Figs.~\ref{fig1}(a) and (b).

Randomness in the $xy$ plane was obtained by randomly walking CsWO nanoparticles from the perfect metasurface.
We selected one nanoparticle in the unit cell and
move it by $\pm 0.02$ nm in $x$ and $y$ directions on condition that 
the nearest-neighbor nanoparticle gap was kept above 0 nm. 
This process was repeated 200,000 times.
Metasurfaces with randomly dispersed nanoparticles obtained experimentally show larger randomness 
in particle position at lower $EC$~\cite{Tani:14}.
The randomness in the $xyz$ direction was obtained by 
first shifting the nanoparticles 
along the $z$ direction randomly from the perfect metasurface 
and then randomly walking in the $xy$ plane as described above. 
The shift along the $z$ direction was sampled from a normal distribution.
The standard deviation of the normal distribution set to
the nearest-neighbor nanoparticle gap distribution of the random metasurface in the $xy$ plane.
For the randomness along the $z$ direction is caused by the non-uniformity of the organic molecular coating,
randomness assumption along the $z$ direction was appropriate.
When we consider a randomness in the shape of CsWO nanoparticles, 
we assume that they are spheroids with axial lengths $a=b$ and $c$ (see the Supporting Information), 
where the long axis is assumed to lie in the $xy$ plane. 
$AR$ defined by $a/c$ was sampled from the normal distribution, 
while the volume is kept unchanged. 
Since CsWO nanoparticles obtained by the breakdown method were nearly spherical ($AR$=1)~\cite{machida2016}, 
we assumed a normal distribution with the mean 1 and the standard deviation 0.15. 
To randomize both shapes and positions, we first change the $AR$ of the nanoparticles randomly 
while keeping their position at the perfect metasurface.
For those nanoparticles that overlap spatially with nearby nanoparticles by the deformation, 
they are moved randomly and repeatedly in $xy$ plane, $\pm$0.02 nm at each step, 
until all overlaps are taken away. 
Then, all nanoparticles are randomly moved by $\pm$0.02 nm in $x$ and $y$ directions, 
for 20,000 times, on condition that the nearest-neighbor nanoparticle gas was kept above 0 nm.

The optical responses of the random metasurface were calculated using the FDTD method. 
We employed SALMON~\cite{NODA2019356} for all calculations.
The grid spacings $\Delta x$, $\Delta y$, $\Delta z$, and $\Delta t$ were set as 0.2 nm and
$3.81\times 10^{-4}$ fs, respectively.
The calculation time was 22.86 fs (60,000 time steps).
Three random metasurfaces were created.
The transmission $T$, reflection $R$, and absorption $A$ for each random metasurface were obtained as follow:
\begin{eqnarray}
T(\lambda)&=&\frac{|\int \bm{E}(x,y,z_1,\lambda)dxdy|^2}{|\int \bm{E}^{(i)}(x,y,z_0,\lambda)dxdy|^2}, \\
R(\lambda)&=&\frac{|\int [\bm{E}(x,y,z_0,\lambda)-\bm{E}^{(i)}(x,y,z_0,\lambda)]dxdy|^2}
{|\int \bm{E}^{(i)}(x,y,z_0,\lambda)dxdy|^2}, \\
A(\lambda)&=&1-T(\lambda)-R(\lambda)
\end{eqnarray}
where $\lambda$ is the wavelength of light and $\bm{E}$ and $\bm{E}^{(i)}$ are the Fourier transformation of 
the electric fields for the random metasurface and 
the incident light, respectively.
We took $z_0=-45.2 $ nm
and $z_1=44.8$ nm
as the observed $xy$ plane to calculate $R$ and $T$, respectively.

From the reflection spectra, 
NIR light reflection performance $ R_{\rm NIR}$ was evaluated as follows:
\begin{eqnarray}
  R_{\rm NIR}=\frac{\int_{\rm NIR}\Phi_{\rm sol}(\lambda)R(\lambda) d\lambda}{\int_{\rm NIR}\Phi_{\rm sol}(\lambda) d\lambda},
\end{eqnarray}
where $\Phi_{\rm sol}$ is the solar spectra~\cite{solar}.
The integration interval is set as [760 nm: 2000 nm].

\section{RESULTS}
\begin{figure}[tb]
  \includegraphics[width=\hsize]{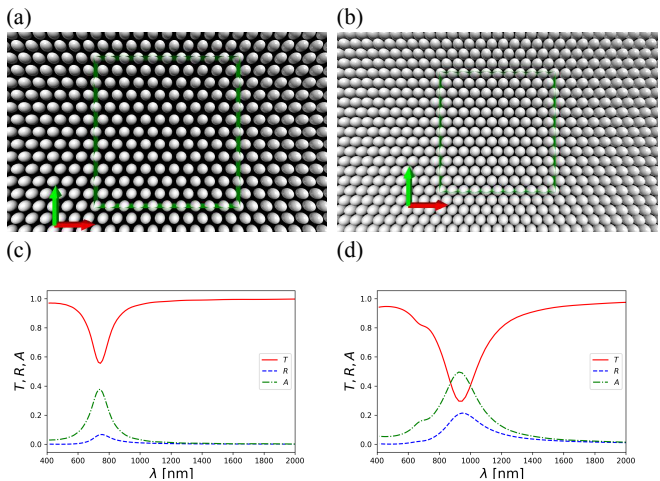}
  \caption{
    Structures of a perfect metasurface and optical spectrum. 
    (a), (b) Structure of a perfect metasurface  at $EC=0.5$ and $EC=0.8$, respectively.
    The white sphere is the CsWO nanoparticles, 
    green frame is the unit cell, 
    red arrow is the $x$ direction, and green arrow is the $y$ direction.
    (c), (d) Transmittance $T$ (red solid line), reflectance $R$ (blue dashed line), 
    and absorptance $A$ (green chain line) at $EC=0.5$ and $EC=0.8$, respectively.
  }
  \label{fig1}
\end{figure}
Figure \ref{fig1} shows the structures and optical spectrum of perfect metasurfaces with $EC=0.5$ and 0.8.
$R_{\rm NIR}$ is calculated as 
0.107 (0.016) for $EC=0.8$ ($EC=0.5$).
As is observed in Ag and ITO metasurface~\cite{Tani:14,Matsui}, 
NIR light reflection is observed in the CsWO metasurface.
As $EC$ increases, the reflectance and absorptance increase, 
with a red-shift in peak due to near-field interaction that occurs as the inter-nanoparticle gap decreases~\cite{Takeuchi}.
Such increasing of $R_{\rm NIR}$ is preferred in heat-shielding glasses.
Based on these results, the effect of randomness is discussed.

\begin{figure}[htb]
  \includegraphics[width=\hsize]{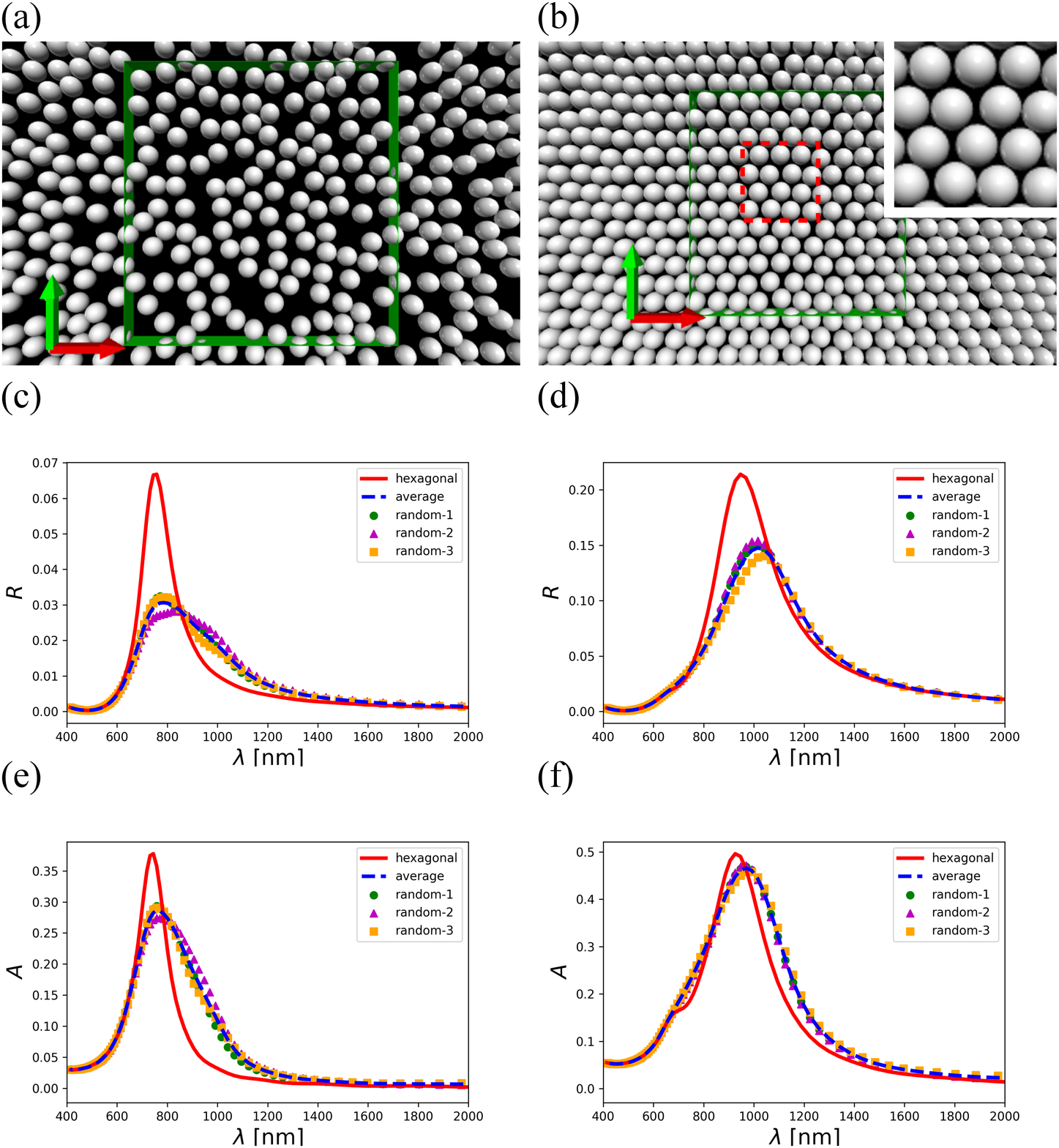}
  \caption{
    Structures and optical spectrum of $xy$ plane random metasurfaces. 
    (a), (c), (e) $EC=0.5$, (b), (d), (f) $EC=0.8$.
    (a), (b) Structures. 
    The inset shows the enlarged red frame region.
    (c), (d) Reflectance $R$ and 
    (e), (f) absorptance $A$.
    The solid red lines and blue dotted lines indicate 
    results of the perfect and average random metasurfaces. 
    The circles, squares, and triangles indicate the results of the three random metasurfaces.
  }
  \label{fig2}
\end{figure}

Figure \ref{fig2} shows the results of the $xy$ plane random metasurfaces.
We prepare three metasurfaces of different random positions, and calculated their optical responses.
A larger randomness of the nanoparticle arrangement appears for lower $EC$ metasurface, 
as seen from Fig.~\ref{fig2}(a) and (b). 
Figure \ref{fig2} (c-f) reveal the reflectance and absorptance of the $xy$ random metasurfaces. 
A comparison with the perfect metasurface shows that the variation in the reflectance 
and absorptance among three random metasurfaces are very small. 
It indicates that the unit cell that contains 120 nanoparticles is sufficiently large to take 
account of the effect of the randomness. 
We will show only an average of three random configurations in later figures.

For $EC=0.5$, the randomness reduces the maximum reflectance to 46\% of that from a perfect metasurface. 
In contrast, for $EC=0.8$, the decrease in maximum reflectance is 69\%,
since the randomness is low compared to that at $EC=0.5$.
However, $R_{\rm NIR}$ is 101\% of the perfect metasurface for $EC=0.5$, 
and 78\% for $EC=0.8$.
In a random metasurface the inter nanoparticle distance can be closer than that in a perfect metasurface.
This enhances the near-field effect, with a red-shift in the reflectance peak.
Thus, $\Phi_{\rm sol}R $ is higher than that of a perfect metasurface in the wide NIR region 
(see Fig.~S3).
Thus, for $EC=0.5$, $R_{\rm NIR}$ is barely changed by the randomness. 

Although the absorptance shows qualitatively the same changes as reflectance, 
the decrease is small.
The absorption is present
in the isolated nanoparticles, 
indicating that the decrease in absorptance due to randomness is small.
In contrast, the reflection originates from the collective effect 
of aligned nanoparticles.
Thus the reflectance rapidly decreases with increasing randomness.

\begin{figure}[tb]
  \includegraphics[width=\hsize]{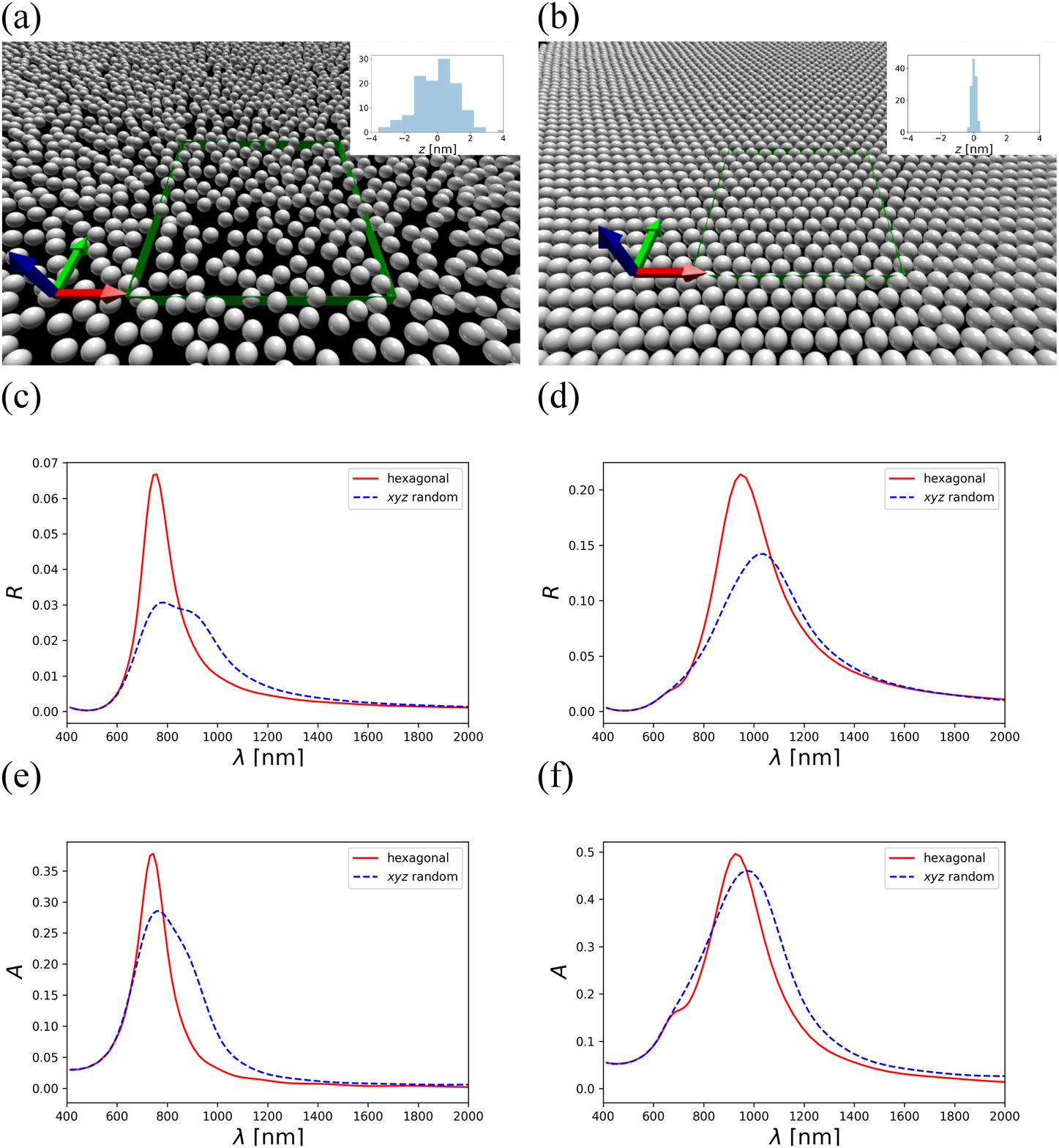}
  \caption{
    Structures and optical spectrum for $xyz$ random metasurface. 
    (a), (c), (e) $EC=0.5$, (b), (d), (f) $EC=0.8$.
    (a), (b) Structures. 
    The inset shows the distribution of the $z$ coordinate of each particle.
    (c), (d) Reflectance $R$ and (e), (f) absorptance $A$.
    We only show the average of the three structures.
  }
  \label{fig3}
\end{figure}

Figure \ref{fig3} reveals the results of $xyz$ random metasurface.
Similar to the $xy$ plane random metasurface, 
$EC=0.5$ has a larger randomness than $EC=0.8$.
Figures \ref{fig3} (a) and (b) show the distribution of $z$ coordinate of the CsWO nanoparticles.
The distribution of the $z$ coordinate is set the same as that of the nearest-neighbor nanoparticle gap in the $xy$ plane random metasurface 
because the non-uniformity along the $z$ direction is 
caused by the non-uniformity of the organic molecule coating.
For $EC=0.5$, since the nanoparticles are randomly dispersed in the $xy$ plane, 
the nearest-neighbor nanoparticle gap in the $xy$ plane is distributed in a wide range. 
Thus, the variation along the $z$ direction is large compared with that for $EC=0.8$.
Figures \ref{fig3} (c-f) show the reflectance and absorptance spectrum of the $xyz$ random metasurface.
No change is observed in $R$ and $A$ with the introduction of randomness along the $z$ direction (comparing Figs.~\ref{fig2} and ~\ref{fig3}).
$R_{\rm NIR}$ is 102\% (76\%) of the perfect metasurface for $EC=0.5$ ($EC=0.8$),
and almost no change is observed with that of the $xy$ plane random metasurface.

\begin{figure}[tb]
  \includegraphics[width=\hsize]{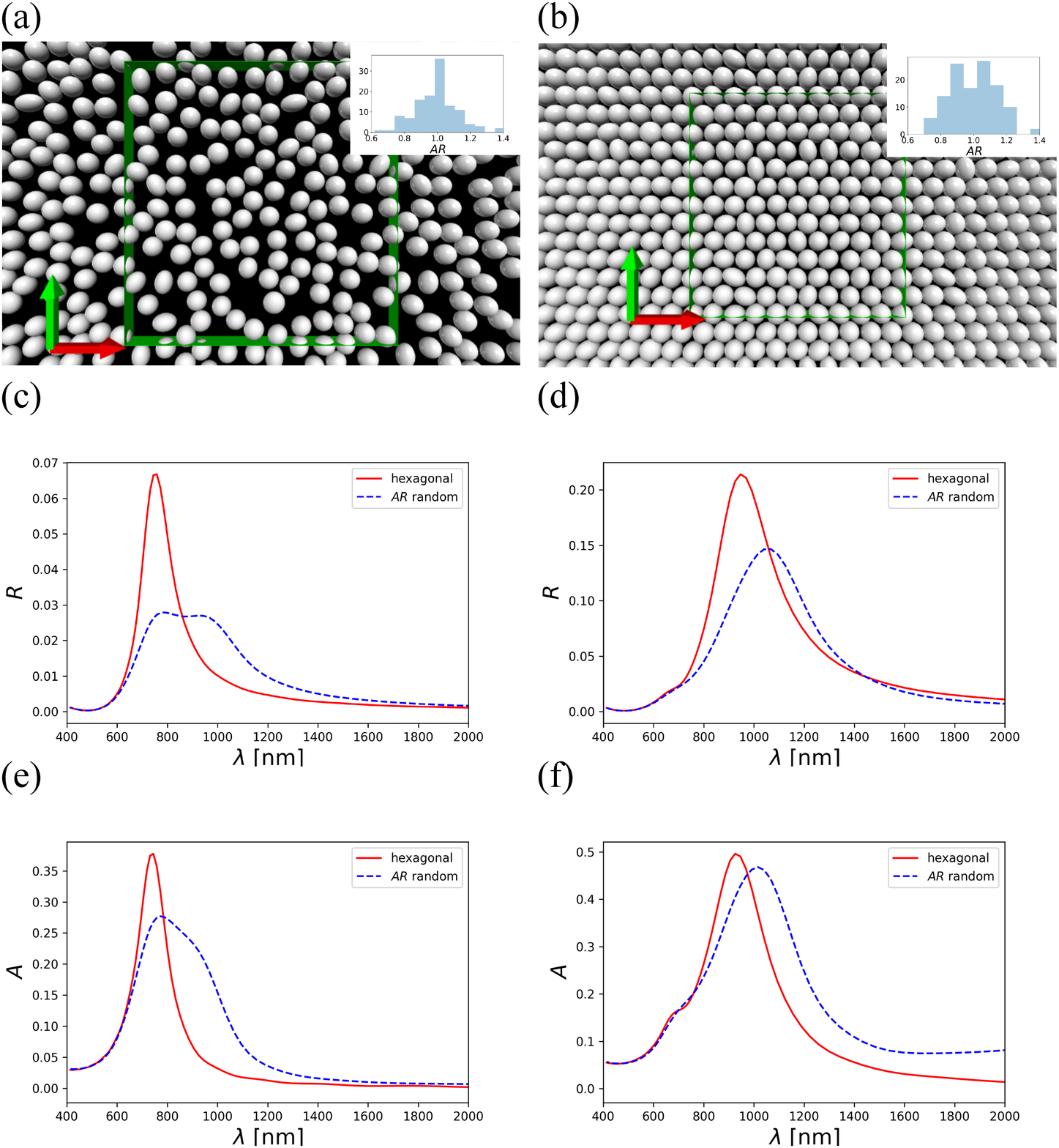}
  \caption{
    Structures and optical spectrum for $xy$ and $AR$ random metasurface.
    (a), (c), (e) $EC=0.5$, (b), (d), (f) $EC =0.8$.
    (a), (b) Structures.
    The inset shows the distribution of $AR$ of each particle.
    (c), (d) Reflectance $R$ and (e), (f) absorptance $A$ of random metasurface.
    We only show the average of the three structures.
    }
  \label{fig4}
\end{figure}
Finally, we discuss the randomness with respect to both
the $xy$ plane and nanoparticle shape.
Figure \ref{fig4} shows the structure, $AR$ distribution, and optical spectrum for
the $xy$ and $AR$ random metasurface.
We set the $AR$ distribution to reflect the CsWO nanoparticles obtained in the experiment~\cite{machida2016}.
Since the nanoparticle changes its spherical shape, there is a significant red shift in the LSPR.
The near-field interaction also contribute to the red shift.
Thus, a red-shift of the peak as well as an enhancement in the long wavelength region appear.
For
$EC=0.5 $, $R_{\rm NIR}$ is increased significantly to 111\% of that of the perfect metasurface.
(For $EC=0.8$, it still decreases significantly to 74\% to that of the perfect metasurface.)
Thus, for low $EC$, the randomness of the nanoparticle shape increases the NIR reflection.

\begin{center}
  \begin{threeparttable}[tb]
    \caption{Paformance ratios.}
    \label{table2}
    \begin{tabular}{cccc}\hline
      $ER$ & Randomness & max$(R_{\rm r})$\tnote{a}/max$(R_{\rm p})$\tnote{b} & $R_{\rm NIR,r}$\tnote{c}/$R_{\rm NIR,p}$\tnote{d} \\
      \hline
      0.5 & $xy$ random & 0.46 & 1.01 \\
      0.5 & $xyz$ random & 0.46 & 1.02 \\
      0.5 & $xy$ and $AR$ random & 0.42 & 1.11 \\
      0.8 & $xy$ random & 0.69 & 0.78 \\
      0.8 & $xyz$ random & 0.66 & 0.76 \\
      0.8 & $xy$ and $AR$ random & 0.68 & 0.74 \\
      \hline
    \end{tabular}
    \begin{tablenotes}
    \item[a] Maximum reflectance of random metasurface
    \item[b] Maximum reflectance of the perfect metasurface
    \item[c] NIR reflection performance of random metasurface
    \item[d] NIR reflection performance of the perfect metasurface
    \end{tablenotes}
  \end{threeparttable}
\end{center}

\section{Conclusions}
This study investigated the NIR reflection performance of 
several random metasurfaces using
the FDTD method.
Preparing metasurfaces with randomness in $xy$ and $xyz$ positions as well as with random spheroidal shapes, 
reflectance, transmittance, and absorptance are calculated for metasurfaces composed of CsWO nanoparticles. 
Taking sufficiently large unit cell that contains 120 nanoparticles, 
we confirmed that randomness is accurately taken into account. 
We made calculations of NIR light reflection performance, $R_{\rm NIR}$, for two effective coverage ($EC$) 
of low value, $EC=0.5$, and high value, $EC=0.8$. 
For low $EC$, $R_{\rm NIR} $ is robust to randomness.
In addition, NIR reflection performance is further improved 
by the randomness of the $AR$ of the nanoparticles.
In contrast, for high $EC$, the NIR reflection performance reduces with increased randomness,
although this value is high for the perfect metasurface.
From an industrial application perspective, low $EC$ has several advantages, 
such as reduced materials and no strict process control because NIR performance is robust against randomness.
However, an increase in the NIR reflection due to the near-field effect cannot be expected.
Therefore, it is necessary to use the materials that induce strong LSPR.
This study can contribute to realizing low-cost, transparent reflective materials 
with new methods, such as material informatics.

%%%%%%%%%%%%%%%%%%%%%%%%%%%%%%%

%SSSSSSSSSSSSSSSSSSSSSSSS
\begin{acknowledgments}
Calculations were carried out at Oakforest-PACS at JCAHPC through 
the Multidisciplinary Cooperative Research Program
in Center for Computational Sciences, University of Tsukuba.
This research was supported by JST-CREST under Grant No. JP-MJCR16N5, 
and also by JSPS KAKENHI Nos. 20H02649 and 20J00449.
\end{acknowledgments}
%SSSSSSSSSSSSSSSSSSSSSSSS

%%% References %%%%%
\bibliography{references}

\end{document}

\title{
  Supporting Information for:
  Infrared-Shielding of Plasmonic Random Metasurface Constructed by Cesium-Doped Tungsten Bronze
}

\author{Tomohiro Yoshida$^{1}$}
\author{Takashi Takeuchi$^{2}$}
\author{Kazuhiro Yabana$^{2}$}

\affiliation{$^{1}$
  Department of Computer-Aided Engineering and Development,
  Sumitomo Metal Mining Co., Ltd., 3-5, Sobiraki-cho, Niihama, Ehime 792-0001, Japan
}

\affiliation{$^{2}$
  Center for Computational Sciences, University of Tsukuba, Tsukuba 305-8577, Japan
}

%%%%%%%%%%%%%%%%%%

\maketitle
\newpage
\begin{figure}[H]
  \includegraphics[width=\hsize]{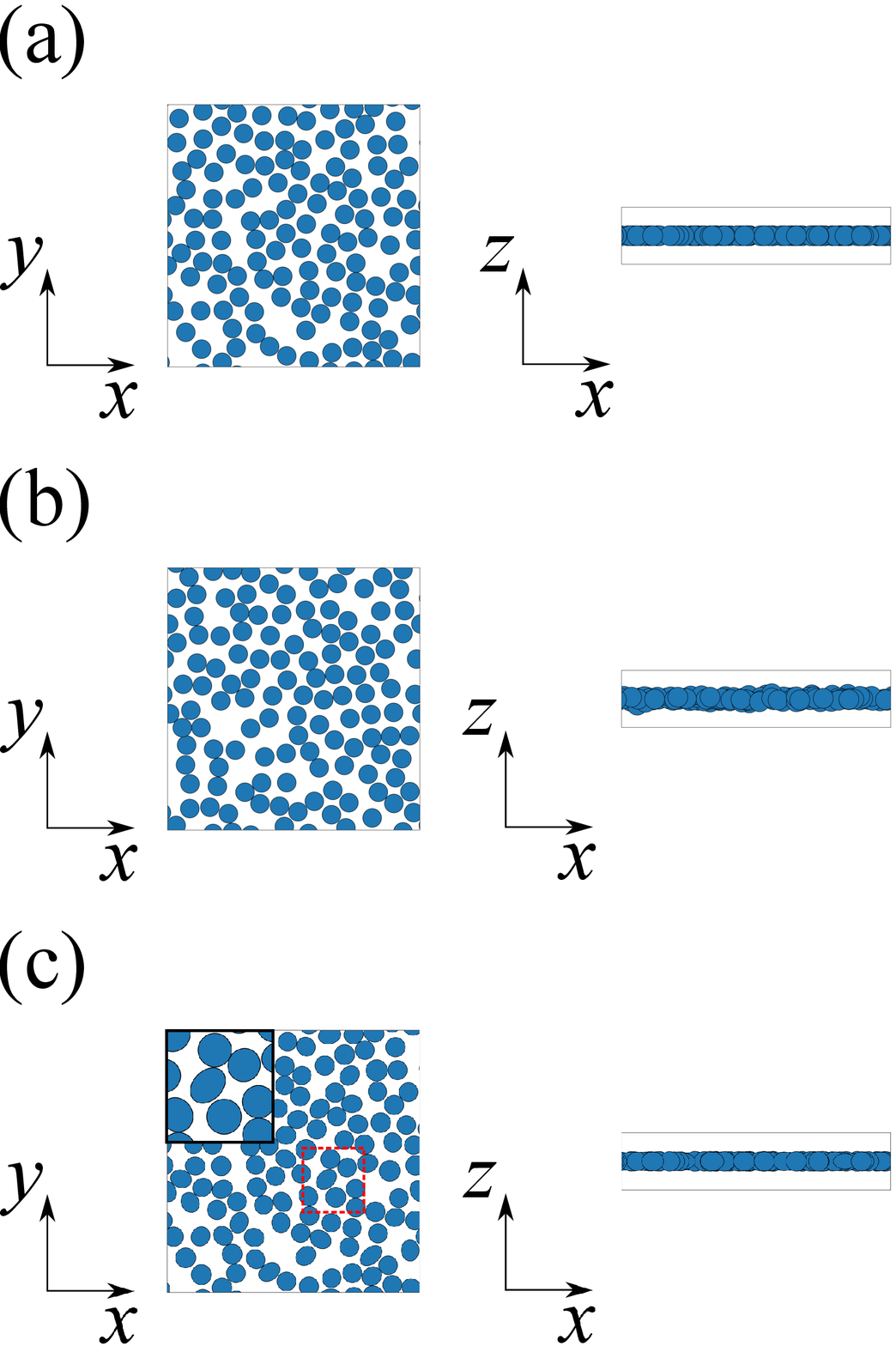}
  \caption{
    Schematic figure of random metasurfaces.
    (a) $xy$ plane random metasurface, (b) $xyz$ random metasurface, 
    and (c) $xy$ plane and nanoparticle shape random metasurface.
    The inset shows the enlarged red frame region.
    }
  \label{S1}
\end{figure}

\begin{figure}[H]
  \includegraphics[width=\hsize]{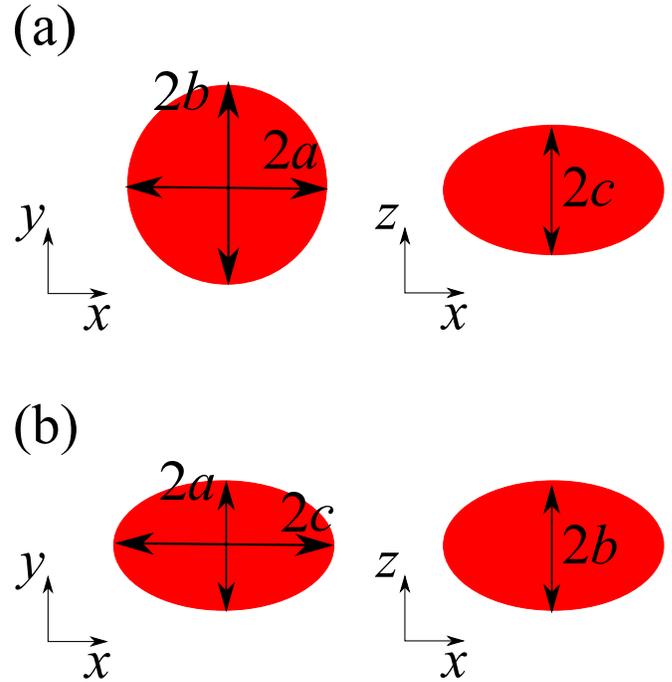}
  \caption{
    Shape of nanoparticles.
    (a) $AR>1$ and (b) $AR<1$.
    }
  \label{S2}
\end{figure}

\begin{figure}[H]
  \includegraphics[width=\hsize]{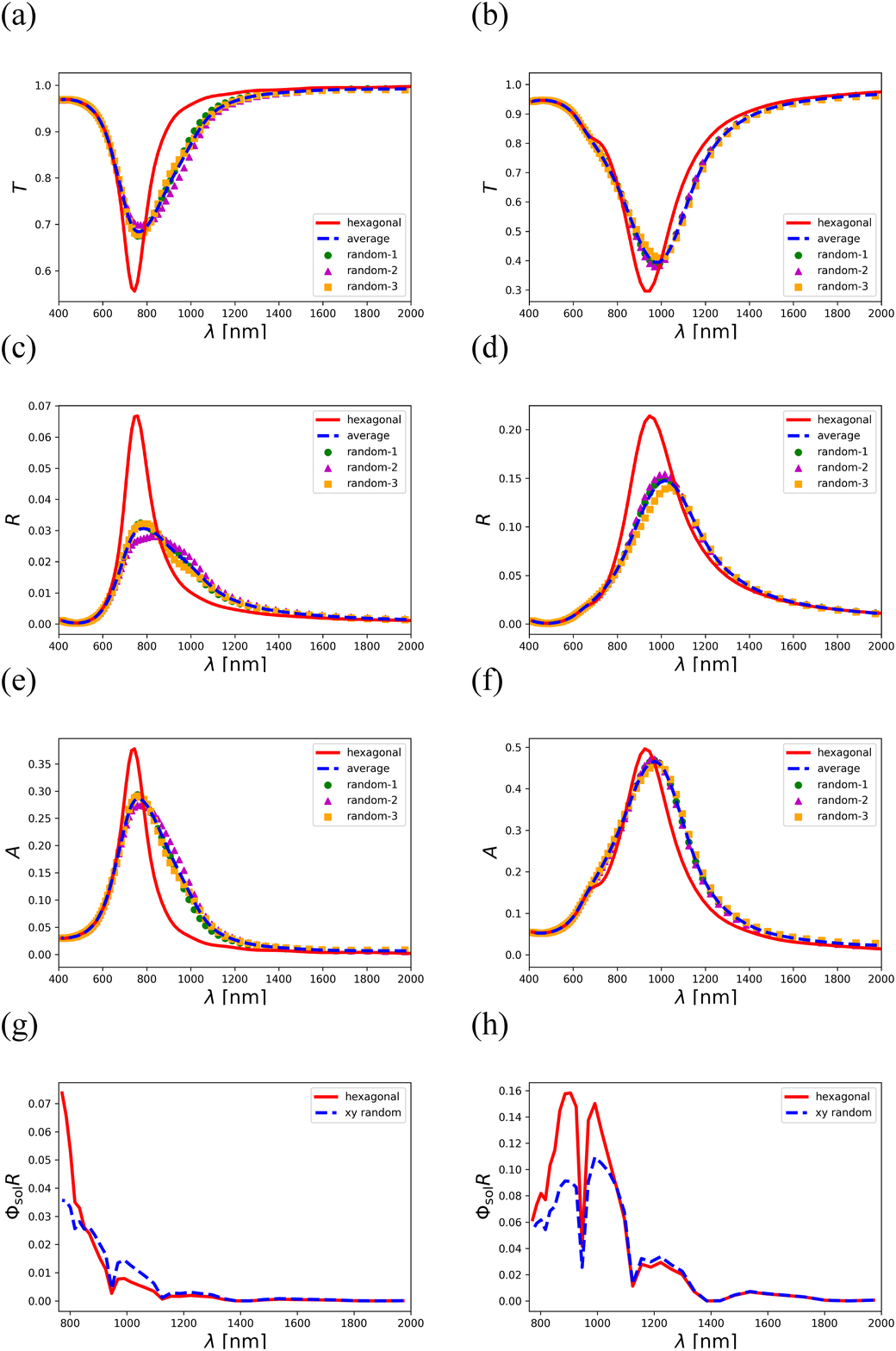}
  \caption{
    Optical spectrum for $xy$ random metasurface.
    (a), (c), (e), (g) $EC=0.5$, (b), (d), (f), (h) $EC =0.8$.
    (a), (b) Transmittance $T$, (c), (d) reflectance $R$, (e), (f) absorptance $A$,
    and (g), (h) $\Phi_{\rm sol}R$ of random metasurface.
    }
  \label{S3}
\end{figure}

\begin{figure}[H]
  \includegraphics[width=\hsize]{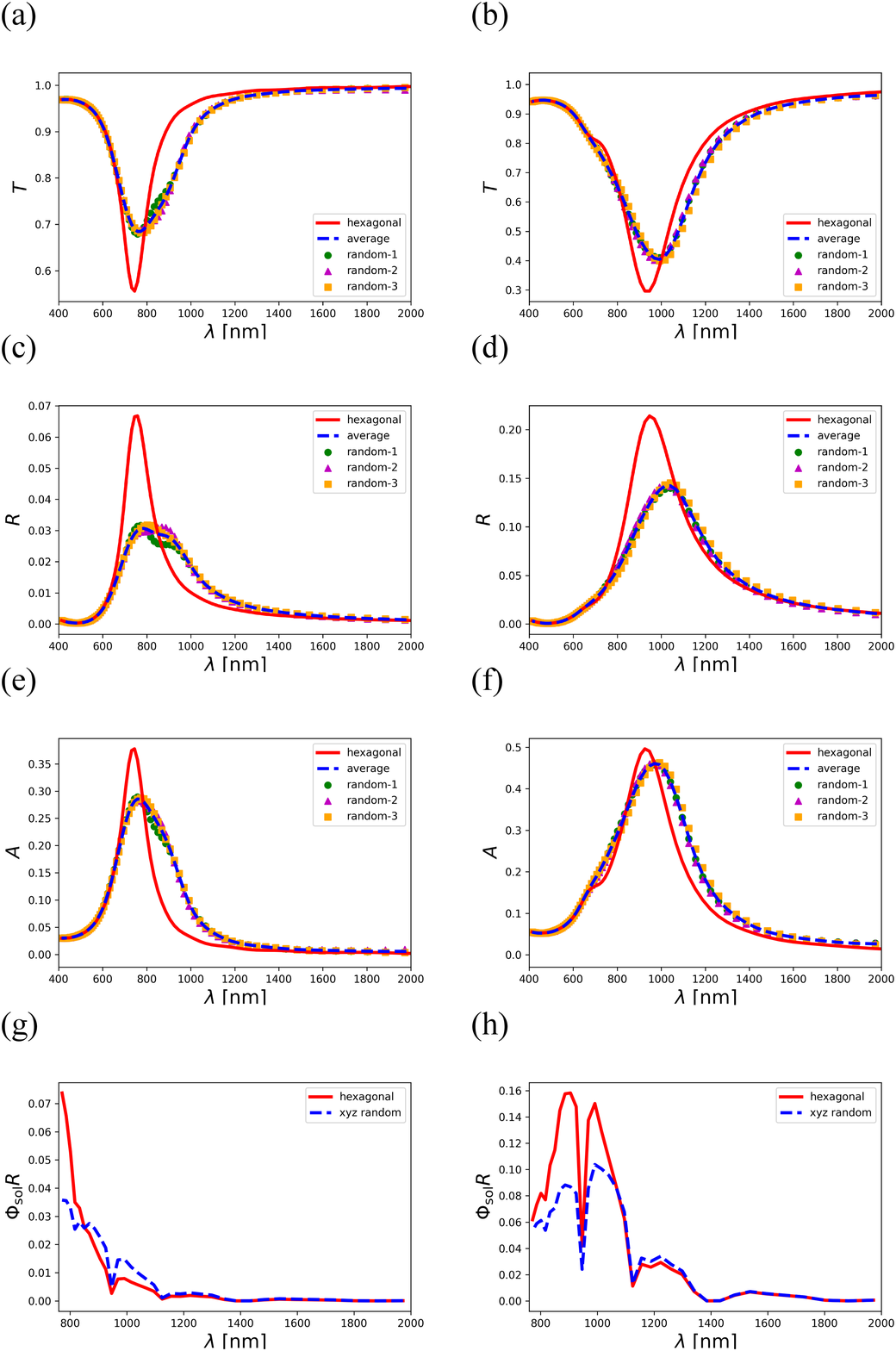}
  \caption{
    Optical spectrum for $xyz$ random metasurface.
    (a), (c), (e), (g) $EC=0.5$, (b), (d), (f), (h) $EC=0.8$.
    (a), (b) Transmittance $T$, (c), (d) reflectance $R$, (e), (f) absorptance $A$,
    and (g), (h) $\Phi_{\rm sol}R$ of random metasurface.
    }
  \label{S4}
\end{figure}

\begin{figure}[h]
  \includegraphics[width=\hsize]{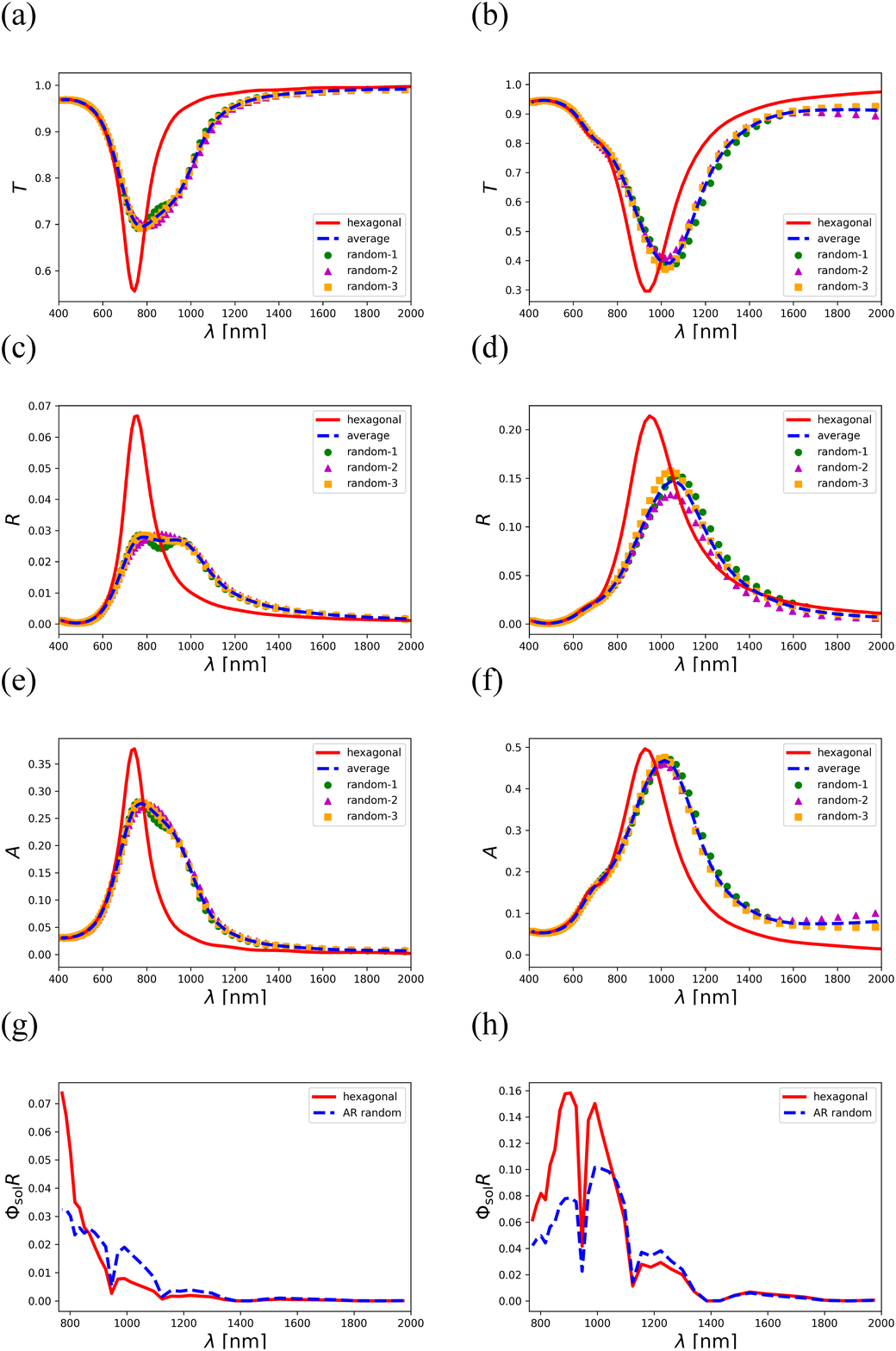}
  \caption{
    Optical spectrum for $xy$ and $AR$ random metasurface.
    (a), (c), (e), (g) $EC=0.5$, (b), (d), (f), (h) $EC=0.8$.
    (a), (b) Transmittance $T$, (c), (d) reflectance $R$, (e), (f) absorptance $A$,
    and (g), (h) $\Phi_{\rm sol}R$ of random metasurface.
    }
  \label{S5}
\end{figure}